\documentclass[prb,superscriptaddress,twocolumn]{revtex4-2}

\usepackage{amsmath}
\usepackage{amsfonts}
\usepackage{amssymb}
\usepackage{txfonts}
\usepackage{bm}
\usepackage{tabularx}
\usepackage{graphicx,color}
\usepackage{ulem}
\usepackage{mathrsfs}

\def\Vec#1{\bm{#1}}
\def\Hc2{H_\mathrm{c2}}
\def\Tc{T_\mathrm{c}}

\begin{document}

\title{
Probing magnetic-field-induced multipolar ordering\\ through field-angle-resolved magnetostriction and thermal expansion in PrIr$_2$Zn$_{20}$
}

\author{Naoki Okamoto}
\affiliation{Department of Physics, Faculty of Science and Engineering, Chuo University, Bunkyo, Tokyo 112-8551, Japan}
\author{Yohei Kono}
\affiliation{Department of Physics, Faculty of Science and Engineering, Chuo University, Bunkyo, Tokyo 112-8551, Japan}
\author{Takahiro Onimaru}
\affiliation{Department of Quantum Matter, Graduate School of Advanced Science and Engineering, Hiroshima University, Higashi-Hiroshima, Hiroshima 739-8530, Japan}
\author{Keisuke T. Matsumoto}
\affiliation{Graduate School of Science and Engineering, Ehime University, Matsuyama, Ehime 790-8577, Japan}
\author{Kazumasa Hattori}
\affiliation{Department of Physics, Tokyo Metropolitan University, 1-1, Minami-osawa, Hachioji, Tokyo 192-0397, Japan}
\author{Shunichiro Kittaka}
\thanks{kittaka@g.ecc.u-tokyo.ac.jp}
\affiliation{Department of Physics, Faculty of Science and Engineering, Chuo University, Bunkyo, Tokyo 112-8551, Japan}
\affiliation{Department of Basic Science, The University of Tokyo, Meguro, Tokyo 153-8902, Japan}

\date{\today}

\begin{abstract}
We performed field-angle-resolved magnetostriction and thermal-expansion measurements on PrIr$_2$Zn$_{20}$, 
a cubic non-Kramers compound exhibiting antiferroquadrupolar order below $T_{\rm Q}=0.125$~K. 
Thermal expansion exhibits two qualitatively different anomalies under magnetic fields applied along the $[001]$ direction, 
providing experimental support for the existence of an intermediate A phase previously reported. 
Furthermore, comparison between the experimental results and theoretical modeling indicates a strong anisotropic coupling of the $O_{20}$ quadrupolar moment,
which plays a key role in stabilizing the A phase. 
These findings demonstrate that multipolar states in non-Kramers systems can be effectively tuned by magnetic-field orientation, 
providing insights into the anisotropic nature of quadrupolar interactions.
\end{abstract}

\maketitle

Strongly correlated electron systems with orbital degrees of freedom provide a fertile ground for realizing exotic multipolar orders beyond conventional magnetic dipoles.
Among these, cubic non-Kramers systems are particularly compelling due to their non-magnetic $\Gamma_3$ doublet ground state, which excludes dipolar contributions and allows quadrupolar and higher-rank multipoles to dominate the low-temperature behavior.
The Pr$T_2X_{20}$ family ($T$: transition metal, $X$: Zn, Al) has emerged as a prototypical platform for investigating such phenomena, exhibiting rich phase diagrams driven by quadrupolar interactions~\cite{Onimaru2016JPSJ, Onimaru2010JPSJ, Onimaru2011PRL, Onimaru2012PRB, Sakai2011JPSJ}.
These compounds crystallize in a cubic structure and host a non-Kramers $\Gamma_3$ doublet arising from the $4f^2$ configuration of Pr$^{3+}$ ions.
The large energy separation from excited crystal electric field states effectively isolates the $\Gamma_3$ doublet, making these systems ideal for studying quadrupolar physics without interference from magnetic dipole moments.

To gain deeper insight into quadrupolar ordering in non-Kramers systems, we focus on PrIr$_2$Zn$_{20}$,
which exhibits an antiferroquadrupolar (AFQ) phase transition at $T_{\rm Q}=0.125$~K and a superconducting transition at $\Tc=0.05$~K in zero magnetic field~\cite{Onimaru2011PRL}.
Specific-heat measurements confirm the nonmagnetic nature of the $\Gamma_3$ ground state, separated from the first excited $\Gamma_4$ state by approximately 30~K~\cite{Onimaru2011PRL}.
Multipolar order has been investigated using a variety of complementary techniques, including neutron scattering~\cite{Iwasa2017PRB}, ultrasound~\cite{Ishii2011JPSJ,Ishii2025PRB}, and $\mu$SR~\cite{Higemoto2012PRB} measurements, whereas NMR measurements are challenging in PrIr$_2$Zn$_{20}$ due to the lack of suitable nuclear moments, in contrast to the case of PrTi$_2$Al$_{20}$~\cite{Taniguchi2019JPSJ}. In particular,
neutron scattering experiments under magnetic fields applied along the $[110]$ direction have identified the ordering vector 
$\Vec{k} = (1/2, 1/2, 1/2)$, consistent with a quadrupolar order of the $v\sim \sqrt{3}(x^2-y^2)$ ($O_{22}$) type~\cite{Iwasa2017PRB}. 
Moreover, PrIr$_2$Zn$_{20}$ exhibits non-Fermi-liquid (NFL) behaviors above $T_{\rm Q}$, 
which are attributed to the formation of a quadrupole Kondo lattice arising from the hybridization between the nonmagnetic quadrupolar degrees of freedom and conduction electrons~\cite{Onimaru2016PRB}.

Recent thermodynamic investigations of PrIr$_2$Zn$_{20}$ have revealed a second anomaly in the specific heat within a narrow magnetic field range along the $[001]$ direction, 
indicating the emergence of a field-induced intermediate phase, referred to as the A phase~\cite{Kittaka2024PRB}. 
This phase appears exclusively in the high-temperature region slightly above $T_{\rm Q}$ and exhibits pronounced anisotropy with respect to the magnetic field orientation. 
Notably, the A phase is absent when the field is applied along the $[110]$ axis, suggesting a strong directional dependence of the underlying order parameter. 
These findings establish PrIr$_2$Zn$_{20}$ as a model system for studying orbital ordering phenomena in non-Kramers doublet systems.

A recent theoretical study by Okanoya and Hattori has proposed a phenomenological interpretation of the A phase based on a Landau free-energy framework incorporating $\Gamma_3$ quadrupolar order parameters $(u,v)$ at $\Vec{k}= (1/2, 1/2, 1/2)$~\cite{Hattori}.
In this model, the A phase arises from internal rotations of the orbital moments associated with the non-Kramers $\Gamma_3$ doublet, driven by the magnetic field. 
In this phase, a collinear $u\sim 2z^2-x^2-y^2$-type quadrupolar order emerges. 
Importantly, the theory accounts for the absence of the A phase under fields applied along $[110]$, leading to strong anisotropy in the phase diagram depending on the field orientation.
This theoretical framework provides a compelling basis for interpreting the field-induced phenomena observed in PrIr$_2$Zn$_{20}$.

In this study, we perform high-resolution field-angle-resolved magnetostriction and thermal-expansion measurements on a single-crystalline sample of PrIr$_2$Zn$_{20}$.
These techniques enable the sensitive detection of anisotropic lattice responses, which reflect the underlying quadrupolar degrees of freedom and their coupling to external magnetic fields.
The experimental results support the theoretical scenario in which the $O_{20}$-type component actively contributes to the stabilization of the A phase.

\begin{figure}
\includegraphics[width=3.4in]{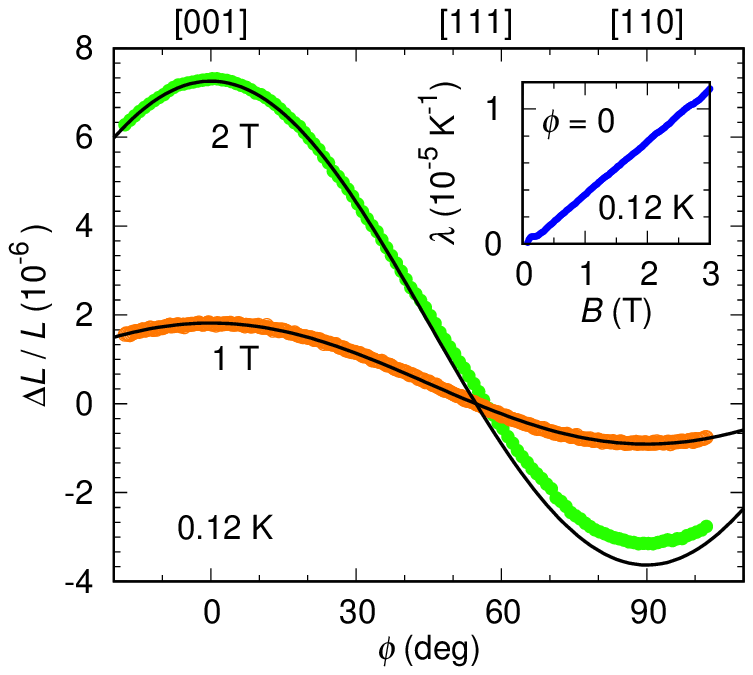}
\caption{
Field-angle $\phi$ dependence of $\Delta L/L=(L-L_0)/L$ at 0.12 K under magnetic fields of 1 and 2~T rotated within the $(1\bar{1}0)$ plane. 
Here, $\phi$ is the magnetic-field angle measured from the $[001]$ axis.
The solid lines show a phenomenological expression for the effective quadrupolar field component, proportional to $B^2(3\cos^2\phi-1)$, which couples to the $O_{20}$ moment (see text for details).
The inset shows the magnetostriction coefficient $\lambda=(\partial L/\partial B)/L$ at 0.12~K for $B \parallel [001]$ (at $\phi=0^\circ$).
}
\label{H}
\end{figure}

The single-crystalline sample used in this study is identical to that employed in previous thermal measurements~\cite{Kittaka2024PRB}.
The orientation of the crystalline axes was determined using X-ray backscattering Laue diffraction (Rigaku, RASCO-BL II), 
which revealed that the measurement axis for the sample length is unintentionally tilted by approximately $8^\circ$ from the $[1\bar{1}0]$ direction toward $[001]$. 
The sample length $L$ is approximately 0.33~mm. 
Further polishing or cutting was avoided to preserve a larger sample volume.
The relative length change, $\Delta L(T,B,\phi)=L(T,B,\phi)-L_0$, as a function of temperature $T$, magnetic field $B$, and field angle $\phi$, was measured using a home-built capacitive dilatometer with a resolution of 0.01~$\AA$~\cite{Kittaka2023PRB,Kittaka2025arXiv},
where $L_0$ is a constant reference length.
The dilatometer was mounted in a dilution refrigerator (Oxford, Kelvinox AST Minisorb), with the $[1\bar{1}0]$ axis approximately aligned along the vertical $z$ direction. 
Magnetic fields were generated using a vector magnet capable of producing up to 7~T horizontally and 3~T vertically. 
The refrigerator was rotated around the $z$ axis using a stepping motor, allowing precise control of the magnetic field orientation in three dimensions.
The data presented here were obtained under magnetic fields precisely aligned within the $(1\bar{1}0)$ plane, nearly perpendicular to the axis 
along which the sample length is measured.
The alignment was optimized based on the field-angle dependence of the magnetostriction of the sample (see Fig.~\ref{H}), achieving an accuracy better than $0.1^\circ$.
This high-precision alignment ensured reliable directional control throughout the measurements.

Magnetostriction and thermal-expansion measurements are widely recognized as effective probes for multipolar degrees of freedom, as they directly couple to lattice distortions and are sensitive to changes in hybridization and valence ~\cite{Worl2019PRB,Worl2022PRR}. 
To ensure the reliability of our magnetostriction measurements, we compared the magnetostriction with previously reported data.
The inset of Fig.~\ref{H} displays the magnetostriction coefficient, defined as $\lambda = (\partial L/\partial B)/L$, measured under a magnetic field applied along the $[001]$ axis.
The coefficient $\lambda$ increases monotonically with increasing field strength, at least up to 3~T.
Although the direction of the measured sample length ($L \parallel [1\bar{1}0]$) differs from that in the previous study ($L \parallel [001]$),
the magnitude of the change agrees well with the previous results~\cite{Worl2019PRB}. 

Figure~\ref{H} plots the relative change in the sample length, $\Delta L$, at 0.12 K, normalized by $L$, as a function of the magnetic field angle $\phi$ measured from the $[001]$ axis. 
The magnetic field is rotated within the $(1\bar{1}0)$ plane, and $L_0=L(0.12\ {\rm K}, 1\ {\rm T}, 54.7^\circ)$. 
Sinusoidal behavior is observed at both 1 and 2~T within the AFQ phase.
When the magnetic field is oriented along the $[111]$ direction ($\phi=54.7^\circ$), the magnetostriction response is nearly insensitive to the field strength. 
This field and field-angle dependence is well reproduced by a phenomenological theory based on a Landau free-energy expansion, as discussed later.
The absence of an asymmetric-angle dependence indicates that the misalignment of the measurement direction is negligible.

\begin{figure}
\includegraphics[width=3.4in]{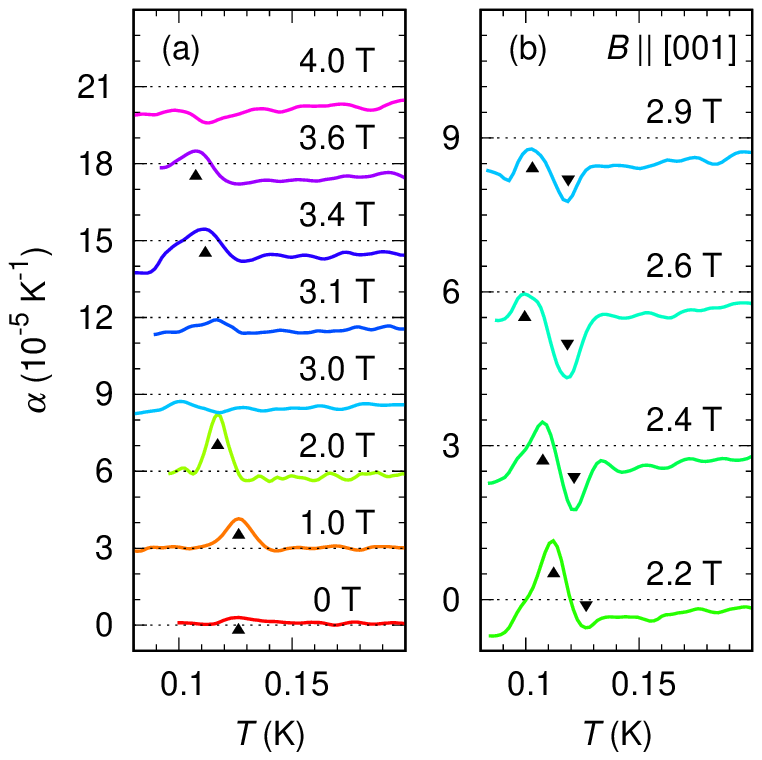}
\caption{
(a) Thermal expansion coefficient $\alpha=(\partial L/\partial T)/L$ measured at several magnetic fields applied parallel to the $[001]$ axis.
(b) Selected field data highlighting the emergence of a double transition.
Each set of data is shifted vertically by $3 \times 10^{-5}$~K$^{-1}$ for clarity. 
Triangles indicate the positions of the anomalies.
}
\label{001}
\end{figure}

Figures~\ref{001}(a) and \ref{001}(b) show the temperature dependence of the thermal expansion coefficient, $\alpha(T) = (\partial L/\partial T)/L$, under magnetic fields from 0 to 4~T applied along the $[001]$ direction.
For fields below 2.2~T, a pronounced contraction is observed at $T_{\rm Q}$ upon cooling, appearing as a prominent peak in $\alpha(T)$ with a positive value.
At 2.4, 2.6, and 2.9~T [see Fig.~\ref{001}(b)], the sample shows a clear expansion, seen as a distinct dip in $\alpha(T)$ with a negative value slightly above $T_{\rm Q}$, alongside a weakened peak at $T_{\rm Q}$, indicating a double transition.
This dip anomaly suggests a field-induced phase transition, consistent with previous specific-heat measurements~\cite{Kittaka2024PRB}.
In contrast, no clear anomaly is seen at 3 and 3.1~T.
At 3.4~T, the data again show a contraction upon cooling through $T_{\rm Q}$.
For fields above 4~T, no distinct anomalies are detected in $\alpha(T)$.

\begin{figure}
\includegraphics[width=3.4in]{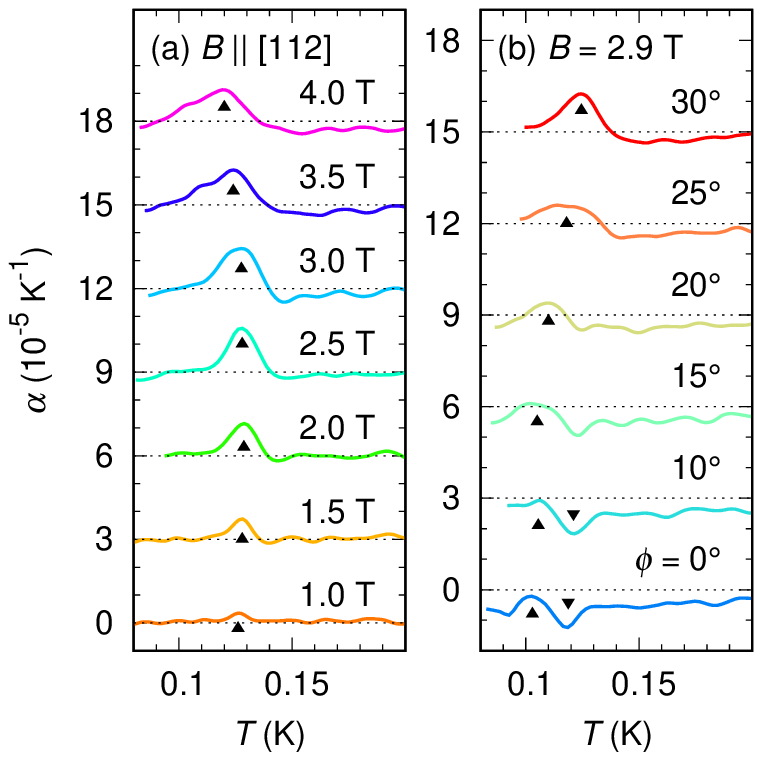}
\caption{
(a) Thermal expansion coefficient $\alpha(T)$ measured at several magnetic fields applied parallel to the $[112]$ axis.
(b) The $\alpha(T)$ data under a magnetic field of 2.9 T rotated in $0^\circ \le \phi \le 30^\circ$.
Each set of data is shifted vertically by $3 \times 10^{-5}$~K$^{-1}$ for clarity.
Triangles indicate the positions of the anomalies.
}
\label{112}
\end{figure}

Figure~\ref{112}(a) shows $\alpha(T)$ under magnetic fields ranging from 1 to 4~T applied along the $[112]$ direction. 
Although previous specific-heat measurements~\cite{Kittaka2024PRB} suggest the occurrence of a double transition at 3 and 3.5~T along the $[112]$ axis, no clear signature of such a transition is observed in the thermal-expansion data.
Instead, the sample exhibits a contraction upon cooling through the transition from the NFL state to the AFQ phase.

To investigate how the double transition observed under a magnetic field along $[001]$ evolves with field orientation, we measured $\alpha(T)$ under a fixed field of 2.9~T while rotating the field from $[001]$ ($\phi=0^\circ$) toward $[112]$ ($\phi=35.3^\circ$). 
As shown in Fig.~\ref{112}(b), the sample length is highly sensitive to the field direction. 
The dip anomaly associated with the higher-temperature transition becomes less distinct as $\phi$ exceeds $20^\circ$.
The origin of this suppression remains unclear.

\begin{figure}
\includegraphics[width=3.4in]{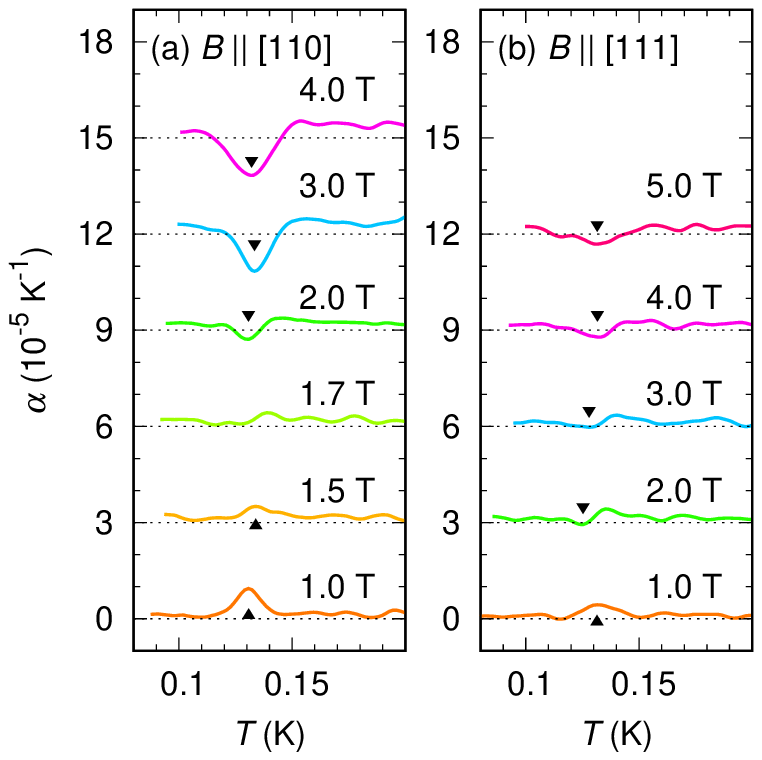}
\caption{
Thermal expansion coefficient $\alpha(T)$ measured at several magnetic fields applied parallel to the (a) $[110]$ and (b) $[111]$ axes.
Each set of data is shifted vertically by $3 \times 10^{-5}$~K$^{-1}$ for clarity.
Triangles indicate the positions of the anomalies.
}
\label{110}
\end{figure}

Figure~\ref{110}(a) displays $\alpha(T)$ measured under magnetic fields applied along the $[110]$ direction.
In the NFL state above $T_{\rm Q}$, $\alpha(T)$ is positive, in contrast to the negative values observed for $B \parallel [001]$ and $[112]$.
As the field increases beyond 1.5~T, the peak anomaly in $\alpha(T)$ gradually diminishes and evolves into a dip with a negative value above 2~T.
This qualitative change may signal a transformation of the order parameter.
Indeed, previous studies have reported an abrupt shift in the specific-heat peak around 2~T~\cite{Kittaka2024PRB} and ultrasonic measurements have revealed anomalies at the same field~\cite{Ishii2011JPSJ}, 
implying the occurrence of another field-induced transition.

\begin{figure}
\includegraphics[width=3.4in]{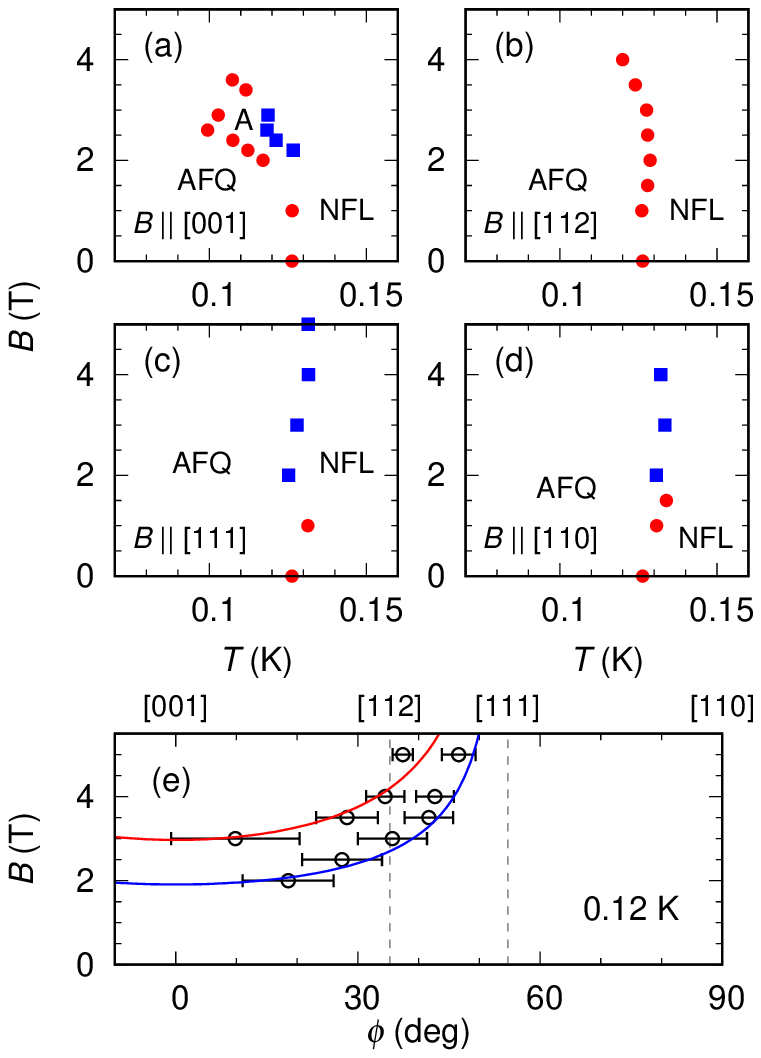}
\caption{
Field-temperature phase diagram of PrIr$_2$Zn$_{20}$ for magnetic fields applied along the (a) $[001]$, (b) $[112]$, (c) $[111]$, and (d) $[110]$ axes. Circles and squares indicate peak and dip anomalies observed in $\alpha(T)$, respectively.
(e) Field-angle $\phi$ dependence of the critical fields obtained from previous specific-heat measurements at 0.12~K~\cite{Kittaka2024PRB}.
The solid lines represent theoretically predictions of $B_{\rm c}(\phi)$ (see text for details) with $B_{\rm c}(0)=1.9$ and $3.0$~T.
}
\label{HT}
\end{figure}

For $B \parallel [111]$, the qualitative behavior of $\alpha(T)$ is likely similar to that for $B \parallel [110]$, 
although the magnitude of the change is relatively small [see Fig.~\ref{110}(b)]. 
The peak and dip anomalies observed in thermal-expansion measurements are summarized in the field-temperature phase diagrams shown in Figs.~\ref{HT}(a)-\ref{HT}(d) for each field orientation.

According to the Ehrenfest relation,
$\partial T_{\rm Q}/\partial p_i=T_{\rm Q}V_{\rm m}\Delta\alpha_i/\Delta c$ should hold for a second-order phase transition, 
where $p_i$ is the uniaxial stress along the $i$ direction, $V_{\rm m}$ is the molar volume, and $\Delta\alpha_i$ and $\Delta c$ are discontinuities in the thermal-expansion coefficient and specific heat at $T_{\rm Q}$, respectively. 
The absence of a clear discontinuity in $\alpha_i$ implies that $T_{\rm Q}$ is relatively insensitive to stress along the $i$ direction. 
In the present case, the measured length change is nearly along the $[1\bar{1}0]$ direction; 
therefore, the disappearance of the dip anomaly in $\alpha(T)$ for $B \parallel [112]$ may suggest that the high-temperature phase transition in this field orientation is insensitive to stress along the $[1\bar{1}0]$.
On the other hand, for $B \parallel [110]$, the sign change in $\Delta\alpha(T)$ was observed, which likely reflects the nature of the uniaxial pressure effect on $T_{\rm Q}$ under magnetic fields.

Based on these results, we examine the possible active order parameter responsible for the field-induced phase transitions in PrIr$_2$Zn$_{20}$. 
In the present study, the sample length change was measured nearly along the $[1\bar{1}0]$ direction.
Within the framework of quadrupole-strain coupling~\cite{Worl2022PRR}, 
the $O_{20}$ uniform quadrupole, which transforms as $2z^2-x^2-y^2$, couples to the strain component $2\varepsilon_{zz}-\varepsilon_{xx}-\varepsilon_{yy}$.
Because the length change along $[1\bar{1}0]$ depends on $\varepsilon_{xx}+\varepsilon_{yy}$, 
the contribution from $O_{20}$ remains finite and is therefore detectable in our measurement.
In contrast, the $O_{22}$ uniform quadrupole, which transforms as $\sqrt{3}(x^2-y^2)$ and couples to $\varepsilon_{xx}-\varepsilon_{yy}$, does not contribute to the length change along $[1\bar{1}0]$, 
since the effects of $\varepsilon_{xx}$ and $\varepsilon_{yy}$ cancel out in this direction.
Consequently, our experimental configuration is sensitive to $O_{20}$ but not to $O_{22}$.
Here, we neglect the possible contribution from the field-induced $T_{xyz}$ octupole moment, 
which may also become active under magnetic fields mainly via the effect of $\varepsilon_{xy}$.

According to the symmetry argument,
the coupling between the magnetic field $\Vec{B}=(B_x,B_y,B_z)$ and the uniform quadrupole moments $O_{20}$ and $O_{22}$ is described by  
\begin{equation}
\mathcal{H}_{\rm Q}=-\beta[ (2B_z^2 - B_x^2 - B_y^2) O_{20} + \sqrt{3}(B_x^2 - B_y^2) O_{22}] \label{eq1}
\end{equation}
where $\beta$ is a coupling constant~\cite{Hattori2014JPSJ}.
When the magnetic field is rotated within the $(1\bar{1}0)$ plane, its components are expressed as 
$\Vec{B}=B(\frac{1}{\sqrt{2}}\sin\phi,\frac{1}{\sqrt{2}}\sin\phi,\cos\phi)$. 
Under this configuration, the field component that couples to $O_{20}$ is given by 
$b_z=2B_z^2-B_x^2-B_y^2=B^2(3\cos^2\phi-1)$.
Notably, according to Eq.~(1), when the magnetic field is oriented along the $[111]$ direction, the coupling term $b_z$ vanishes.
This implies that $O_{20}$ does not couple to the field in this orientation. 
As demonstrated by the solid lines in Fig.~\ref{H}, this theoretical model reproduces the field-angle dependence of the magnetostriction data well,
including the weak magnetostriction response observed near the $[111]$ direction. 
This agreement reinforces the interpretation that quadrupole moments play important role in PrIr$_2$Zn$_{20}$.

To further validate Eq.~(\ref{eq1}), 
we analyzed the angular dependence of the critical field $B_{\rm c}(\phi)$
associated with the double transition observed in previous specific-heat measurements~\cite{Kittaka2024PRB}.
As shown in Fig.~\ref{HT}(e), the experimental data were fitted using the function 
$B_{\rm c}(\phi) =\sqrt{2}B_{\rm c}(0)/\sqrt{3\cos^2\phi-1}$,
which originates from the effective field component $b_z$ that couples to $O_{20}$. 
The good agreement between theory and experiment supports the presence of a finite expectation value of $O_{20}$ and highlights the anisotropic nature of the field-induced phase transitions in PrIr$_2$Zn$_{20}$.

If the sample expansion observed in the NFL state for $B \parallel [001]$, i.e., a negative value in $\alpha(T)$ above $T_{\rm Q}$, corresponds to a positive growth of $O_{20}$, 
then the contraction observed during the transitions from the NFL state to the AFQ phase, and 
from the A phase to the low-temperature AFQ phase, suggests a suppression of $O_{20}$ in the AFQ phase. 
In contrast, the expansion observed during the transition from the NFL state to the A phase implies a further development of $O_{20}$ in the A phase.
These observations reflect underlying order parameter changes.
It is important to note that the $\Gamma_3$ quadrupolar order parameter $(u,v)$ at $\Vec{k}=(1/2,1/2,1/2)$ can couple with the uniform quadrupoles $O_{20}$ and $O_{22}$ as 
\begin{align}
\mathcal{H}'_{\rm Q}=-\gamma[ O_{20}(u^2-v^2) + 2O_{22}uv], 
\end{align}
where $\gamma>0$ is the coupling constant~\cite{Hattori}. 
This readily indicates that an increase in $u^2$ leads to a positive growth of the uniform $O_{20}$, 
whereas an increase in $v^2$ results in its suppression.
Although the previous neutron-scattering experiments under a magnetic field along the $[110]$ direction have suggested that 
the $O_{22}$-type component $v$ is the dominant order parameter in the AFQ phase~\cite{Iwasa2017PRB}, 
the present magnetostriction results indicate that the $O_{20}$-type $u$ also plays a significant role, particularly in the intermediate A phase. 
These findings validate key aspects of recent theoretical models, deepening our understanding of multipolar ordering and its directional sensitivity in non-Kramers doublet systems.
While the transformation of the $\alpha(T)$ peak into a dip above 2 T at $T_{\rm Q}$ for $B \parallel [110]$ remains beyond the scope of current theoretical frameworks,
it opens a promising avenue for future theoretical development and may yield new insights into anisotropic multipolar interactions.
In particular, the emergence of field-induced phases may be partially driven by the field-induced $T_{xyz}$ octupole moment,
whose contribution was not considered in the present analysis.

This study demonstrates that field-angle-resolved magnetostriction and thermal-expansion measurements provide an effective approach for probing multipolar order and its anisotropic lattice responses. 
Combined with other thermodynamic measurements developed for rotating magnetic fields~\cite{Onimaru2004JPSJ,Sakakibara2016RPP,Kittaka2018JPSJ,Kittaka2020JPSJ,Kittaka2021JPSJ,Kittaka2024PRB,Yuasa2025PRB}, 
these efforts highlight the critical role of precise directional tuning in advancing the understanding of complex quantum materials.

In summary, we have performed high-resolution thermal-expansion and magnetostriction measurements to investigate magnetic-field-induced multipolar ordering in PrIr$_2$Zn$_{20}$.
The field-angle-resolved data indicate that the $O_{20}$-type quadrupolar moment serves as a key order parameter in the intermediate A phase.
These findings underscore the rich multipolar degrees of freedom inherent in PrIr$_2$Zn$_{20}$
and establish field-angle-resolved magnetostriction as a powerful technique for probing quadrupole order parameters.

\begin{acknowledgments}
This work was supported by JST FOREST Program (JPMJFR246O) and JSPS KAKENHI (JP23H04868, JP23H04869, JP23H04870, JP23H01128, and JP24K00574).
\end{acknowledgments}

\bibliography{ref_PrIr2Zn20.bib}

\end{document}